\documentclass[showpacs, preprintnumbers, amsmath, amssymb, floatfix, twocolumn]{revtex4-1}

\usepackage{graphicx}
\usepackage{dcolumn}
\usepackage{bm}

\begin{document}

\title{Anomalous Transport Properties of Re${_3}$Ge${_7}$}

\author{Anja Rabus and Eundeok Mun}
\affiliation{Department of Physics, Simon Fraser University, 8888 University Drive, Burnaby, BC, V5A 1S6, Canada}


\date{\today}
\begin{abstract}

Single crystals of intermetallic Re${_3}$Ge${_7}$ were grown and characterized by measuring magnetization, electrical resistivity, Hall coefficient, and specific heat. Magnetization measurements show the material is weakly diamagnetic. A phase transition is indicated by a kink in magnetic susceptibility at $T_{c}$~=~58.5~K and is confirmed by a $\lambda$-like anomaly in specific heat. In zero-field, the temperature dependence of electrical resistivity $\rho(T)$ follows a typical metallic behavior above $T_c$ and sharply increases below $T_c$, showing a metal-to-insulator-like transition. When a magnetic field is applied, strong effects on the transport properties are observed. The temperature dependence of magnetoresistivity $\Delta\rho$ = $\rho (T, H)$ - $\rho (T, H=0)$ develops a maximum around 30 K, deviating from ordinary metallic behavior. Analysis of the Hall coefficient measurements indicates that the carrier density is 0.04 per formula unit at 300 K and drops by two orders of magnitude below $T_c$. The effective mass of charge carriers is inferred from the analysis of the Shubnikov-de Haas quantum oscillations to be close to the bare electron mass.

\end{abstract}


\maketitle


\section{Introduction}

Recently the study of materials with a nontrivial topology of band structure has become an interesting topic, leading to the discovery of new classes of materials such as topological insulators and Dirac, nodal line, and Weyl semimetals.\cite{Hasan2010, Yan2017, Armitage2018} The strong spin-orbit coupling creates a band inversion and induces nontrivial band topology, where the surface state is protected by time reversal symmetry and inversion symmetry. Recently, many binary and ternary compounds with heavy metals have been synthesized and revisited to explore aspects of topologically nontrivial properties. In addition, consequences of spin-orbit coupling appear to be manifested in correlated electron materials such as oxides, superconductors, and topological insulators.\cite{Hasan2010, Yan2017, Armitage2018, Kim2008, Jackeli2009, Singh2010, Ng2000, Qi2011} Because of the atomic structure, topological materials are more likely to exist in compounds with high atomic number.

The rhenium (Re)-based compounds are difficult to synthesize owing to the element's high melting temperature (3186 $^{o}$C) and low solubility into solid solution.\cite{Okamoto2000} As a result, there are only limited binary phase diagrams available and the physical properties of many Re-based compounds remain unexplored. In the Re-Ge binary system, only one binary compound Re$_{3}$Ge$_{7}$ has been reported, which decomposes peritectically at 1132 $^{o}$C.\cite{Okamoto2000} Although the eutectic region is very narrow in the Re-Ge binary phase, we were able to grow single crystals of Re$_{3}$Ge$_{7}$ out of Ge-rich melt. The crystal structure of Re$_{3}$Ge$_{7}$ has been characterized prior to this study as base-centered orthorhombic ($C$mcm, No. 63), with lattice parameters \textit{a}~=~3.2270~\AA, \textit{b}~=~9.0450~\AA, \textit{c}~=~21.9560~\AA.\cite{Siegrist1983} However, there is no investigation of thermodynamic and transport properties of this material. Here we report the thermodynamic and transport properties of single crystals of Re$_{3}$Ge$_{7}$. The observed electronic and magnetic properties characterized by measuring magnetization, electrical resistivity, Hall resistivity, and specific heat are consistent with it being a diamagnetic, intermetallic compound, but with a phase transition at a critical temperature $T_{c} = 58.5$~K.

\section{Experimentals}

\begin{figure}
\centering 
\includegraphics[width=1\linewidth]{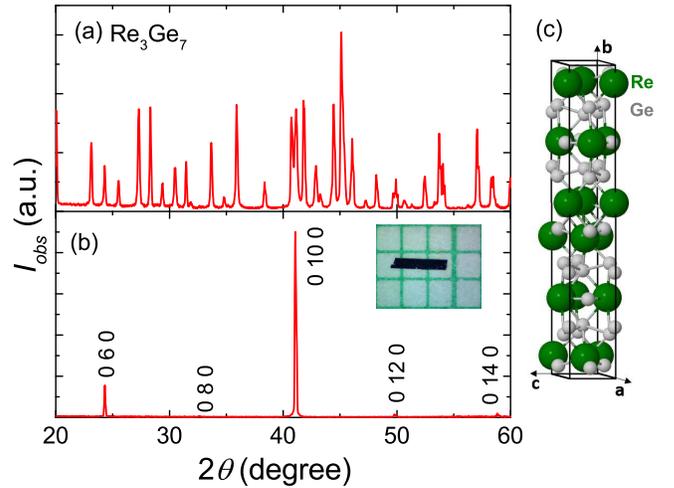}
\caption{X-ray diffraction patterns of Re$_{3}$Ge$_{7}$ on powdered sample (a) and on a single crystal in (0, $k$, 0) orientation (b). Inset shows a photograph of as-grown sample on mm grid. (c) Crystal structure of the Re$_{3}$Ge$_{7}$.}
\label{FIG1}
\end{figure}

Single crystals of Re${_3}$Ge${_7}$ were grown out of Ge-rich melt.\cite{Canfield1992} The original constituents with composition Re$_{0.03}$Ge$_{0.97}$ were placed in an almunia crucible and sealed in a quartz tube under partial Argon pressure. Several batches were heated to $1225^ {\circ}$C, then cooled over varying time periods down to $950^{\circ}$C. The crystals generally grow with a needle-like or ribbon-like morphology, with the longest dimension along the crystallographic $a$-axis, as shown in the inset of Fig. \ref{FIG1}. The crystal structure was determined by powder X-ray diffraction (XRD) measurements in a Rigaku Miniflex diffractometer. Figures \ref{FIG1} (a) and (b) show the powder XRD pattern and single crystal in (0, $k$, 0) orientation, respectively. The detected peaks in the powder XRD pattern can be indexed using an orthorhombic ($C$mcm) structure with lattice parameters \textit{a} = 3.22645 \AA, \textit{b} = 21.95558 \AA, \textit{c} = 9.03557 \AA, which is consistent with the literature values \cite{Siegrist1983}. The crystal structure of Re${_3}$Ge${_7}$ is shown in Fig.~\ref{FIG1}~(c).

The magnetization was measured by using a Quantum Design Superconducting Quantum Interference Device (QD SQUID) magnetometer down to 1.8 K and up to 70 kOe. Because of the small diamagnetic response of Re$_{3}$Ge$_{7}$, multiple pieces of small single crystals (each with mass $\sim$2 mg) were placed into a gel-capsule to measure magnetization. Thus, the magnetization measurement for Re$_{3}$Ge$_{7}$ can be regarded as a polycrystalline average. Specific heat was measured by the relaxation method in a QD Physical Property Measurement System (PPMS). The electrical resistivity down to 1.8 K and up to 90 kOe was measured using a four-probe technique with the current flowing along the $a$-axis ($I \parallel a$) in a QD PPMS. In addition, the resistivity measurement was performed at $T$ = 0.5 K for 50 kOe $< H <$ 90 kOe to measure quantum oscillations. The magnetic field dependence of the resistivity was measured by switching polarity of magnetic field. Hall resistivity measurements were performed in a four-wire geometry, where the magnetic field directions were also reversed to remove magnetoresistance effects due to voltage wire misalignments.

\section{Results}

\begin{figure}
\centering 
\includegraphics[width=1\linewidth]{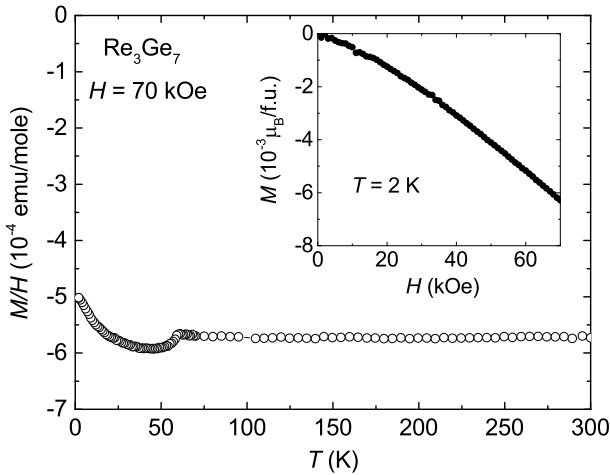}
\caption{Temperature dependence of magnetic susceptibility $M/H$ at $H$ = 70 kOe. Inset shows magnetization isotherm at $T$ = 2 K.}
\label{FIG2}
\end{figure}

The magnetization isotherm at $T$ = 2 K decreases quasi-linearly with increasing magnetic fields as shown in the inset of Fig. \ref{FIG2}, which is consistent with Re$_{3}$Ge$_{7}$ being a weakly diamagnetic compound. Figure \ref{FIG2} shows the temperature-dependent magnetic susceptibility, $M/H$, of Re$_3$Ge$_7$. The $M/H$ is basically temperature-independent at high temperatures, however a sharp slope change is detected at $T_{c}$ = 58.5 K as a signature of phase transition. $T_{c}$ is determined by the peak position in $d\chi(T)/dT$. Below $T_{c}$, a small upturn in the slope is likely due to the presence of paramagnetic impurities.

\begin{figure}
\centering 
\includegraphics[width=1\linewidth]{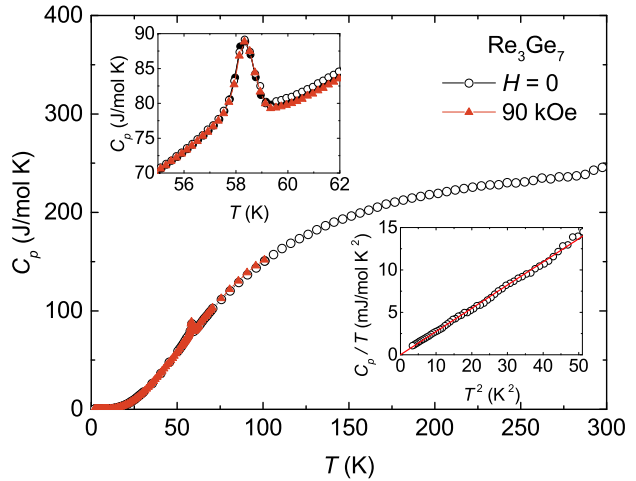}
\caption{Specific heat curves for $H$ = 0 and 90 kOe. Upper inset shows the expanded plot around the $T_{c}$. Solid and open circles correspond to data taken in warming and subsequent cooling temperature, respectively, in zero field. Solid triangles correspond to data taken in warming at $H$~=~90~kOe. Lower inset shows $C_{p}/T$ vs. $T^{2}$.}
\label{FIG3}
\end{figure}

The temperature-dependent specific heat curves of Re$_3$Ge$_7$ for $H$ = 0 and 90 kOe are shown in Fig. \ref{FIG3}. In zero field, the specific heat at high temperatures approaches a value of $\sim$ 250 J/mol K, which is close to the classical Dulong-Petit limit. When cooling the sample from 300 K, the specific heat curve follows the behavior of a typical metal with the exception of a distinct peak structure at $T_{c}$ = 58.4 K (upper inset). The observed $\lambda$-like anomaly clearly indicates a phase transition at $T_{c}$, which is consistent with the magnetic susceptibility result. Note that no thermal hysteresis upon warming and cooling the sample is observed at $T_{c}$. Specific heat measured at 90~kOe applied along $H \parallel b$ follows identical behavior to the zero-field curve. At low temperatures, the specific heat curve can be described by $C_{p}(T) = \gamma T + \beta T^{3}$, where the linear and cubed terms are the electronic and phonon contributions, respectively. The electronic specific heat coefficient $\gamma$ and Debye temperature $\Theta_{D}$ are estimated from the $C(T)/T$ versus $T^{2}$ plot as shown in the lower inset of Fig. \ref{FIG3}. The obtained $\gamma$ is $\sim$ 4 $\mu$J/mol K$^{2}$, which is much smaller than that of Cu \cite{Pobell1996}. The negligibly small $\gamma$ value indicates either a small effective mass of charge carriers or small carrier density of Re$_3$Ge$_7$ compound. The Debye temperature is estimated to be $\Theta_{D}$~$\sim$~190 K.

\begin{figure}
\centering 
\includegraphics[width=1\linewidth]{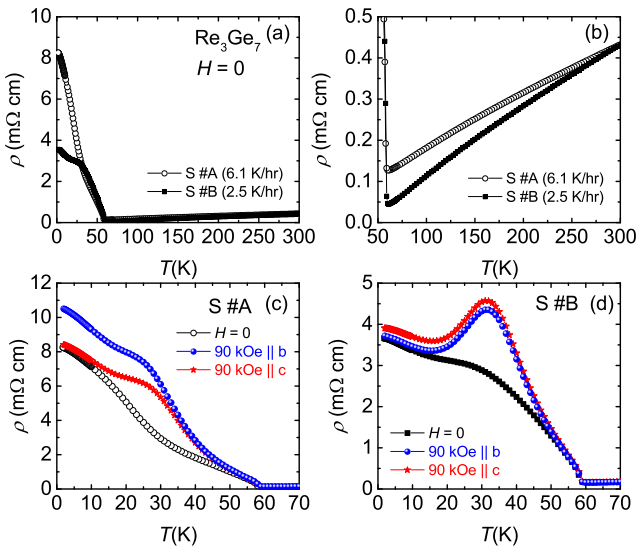}
\caption{  (a) Zero-field resistivity for different growth conditions. Sample A and sample B represent samples grown with a cooling rate of 6.1$^{\circ}$C/hr and 2.5$^{\circ}$C/hr, respectively. (b) Zero-field resistivity above 50 K. (c) Resistivity curves for Sample~A at 90 kOe for $H \parallel b$ and $H \parallel c$. (d) Resistivity curves for Sample B at 90 kOe for $H \parallel b$ and $H \parallel c$.}
\label{FIG4}
\end{figure}

The electrical resistivity of multiple samples, grown under slightly different conditions, has been measured. Figure \ref{FIG4} (a) shows the temperature dependence of the electrical resistivity $\rho(T)$ of two representative Re$_{3}$Ge$_{7}$ samples, where Sample A and Sample B were grown by cooling at rates 6.1$^{\circ}$C/hr and 2.5$^{\circ}$C/hr, respectively. The $\rho(T)$ curves for both samples follow a typical metallic behavior at high temperatures and show a steep increase below $T_c$ = 58.5 K, indicating a metal-to-insulator-like transition. The $T_{c}$ is determined from the peak position in $d\rho(T)/dT$. At high temperatures both $\rho(T)$ curves decrease with decreasing temperature in a nearly linear fashion with slightly different slope as shown in Fig. \ref{FIG4}~(b), where the resistivity curve of Sample A is normalized to the resistivity value of Sample B at 300 K. Whereas the resistivity curve of Sample A increases smoothly as the temperature is decreased from $T_c$ to 1.8 K, the resistivity curve of Sample B plateaus around 30 K, then continues to increase at lower temperatures. Note that, based on the resistivity measurements of multiple samples with both needle-like and ribbon-like morphology, the resistivity at high temperatures appears to display metallic behavior with slightly different slope. However, the resistivity below $T_{c}$ is sensitive to the growth conditions, following the behavior of either Sample A or Sample B.

Figures \ref{FIG4} (c) and \ref{FIG4} (d) show the resistivity curves for both Sample A and B, respectively, measured in $H$~=~90~kOe for $H \parallel b$ and $H \parallel c$. In both samples, the resistivity shows an anomalous temperature dependence below $T_c$, with the maximum variation in resistivity around 30~K. For Sample B, $\rho(T)$ at 90 kOe in both orientations produces a distinct maximum at around 30~K, whereas for Sample A, $\rho(T)$ data show a broad shoulder around 30~K. The resistivity of Sample~B appears not to be sensitive to the orientation of the magnetic field, displaying no obvious anisotropy between $H \parallel b$ and $H \parallel c$. For Sample A, the absolute value of resistivity for $H \parallel b$ is slightly larger than that for $H \parallel c$ below $\sim$ 30 K, but the curves follow the same trend nevertheless.

\begin{figure}
\centering 
\includegraphics[width=1\linewidth]{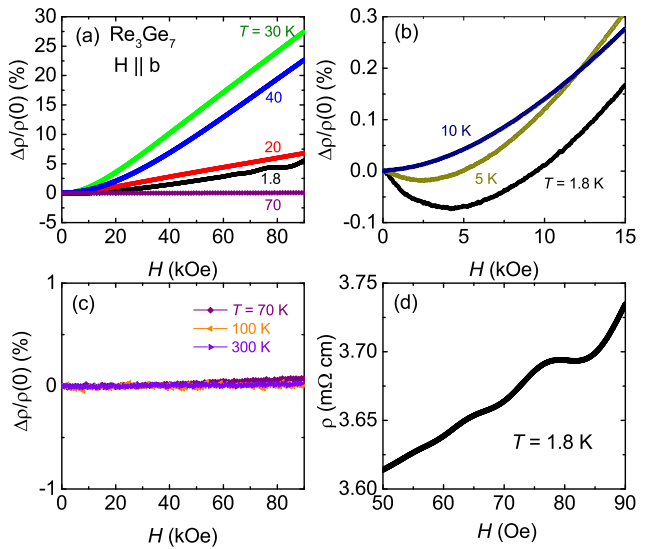}
\caption{(a) Magnetic field-dependent resistivity $\Delta\rho/\rho(0)$ curves for Sample~B at selected temperatures. (b) $\Delta\rho/\rho(0)$ curves at $T$ = 1.8, 5, and 10 K. (c) $\Delta\rho/\rho(0)$ curves at $T$ = 70, 100, and 300 K. (d) Resistivity at 1.8 K for $H >$ 50 kOe.}
\label{FIG5}
\end{figure}

The magnetic field-dependent resistivity (magnetoresistance, MR) curves at several fixed temperatures for Sample B are shown in Fig. \ref{FIG5} (a), where the curves are plotted by $\Delta\rho / \rho(0)$ $\times$ 100 = [$\rho(H) - \rho(0)] / \rho(0)$ $\times$ 100. At 1.8 K the MR is negative in low fields and changes to positive above 10 kOe as shown in Fig. \ref{FIG5} (b). The negative MR in the low field regime is observed below 10 K. Since there are no local moment bearing atoms in Re$_{3}$Ge$_{7}$, the origin of the negative MR in low fields cannot be related to the magnetism. With increasing temperature, the magnitude of MR first increases in the temperature range 1.8$-$30 K and then decreases again up to 70 K, producing the maximum variation of MR around 30 K. This result is consistent with the temperature- dependent resistivity curves at $H$ = 90 kOe, which indicate a peak structure around 30 K (Fig. \ref{FIG4}). The MR curves follow a quadratic field dependence $\Delta\rho/\rho(0) \propto H^{2}$ in the low field regime and change to a linear field dependence $\Delta\rho/\rho(0) \propto H$ at high fields. It should be noted that the MR is negligible above $T_{c}$ as shown in Fig. \ref{FIG5} (c), where the MR varies much less than 1 \% up to 90 kOe. Figure Fig. \ref{FIG5} (d) shows the MR curve at $T$ = 1.8 K above 50~kOe. Pronounced Shubnikov de-Haas (SdH) oscillations are observed in MR at $T <$ 5 K, indicating the high quality of the samples. Note that SdH oscillations are clearly detected for all samples measured (not shown here).

\begin{figure}
\centering 
\includegraphics[width=1\linewidth]{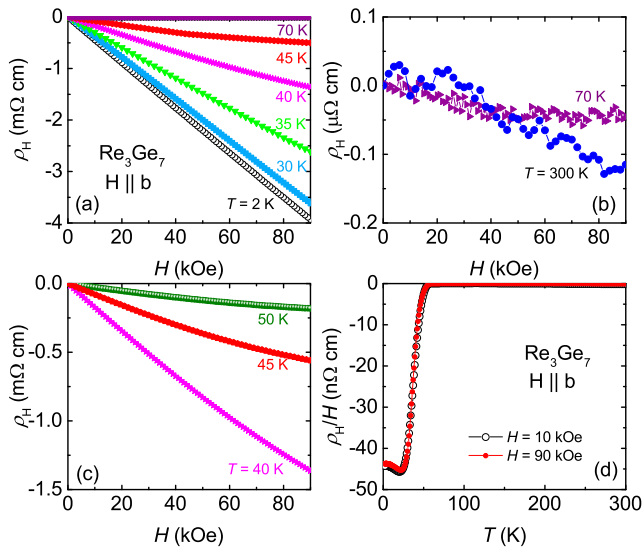}
\caption{(a) Hall Resistivity $\rho_{H}$ curves at selected temperatures. (b) $\rho_{H}$ at $T$ = 70 and 300 K. (c) $\rho_{H}$ at $T$ = 40, 45, and 50 K. (d) Temperature dependence of the Hall coefficient $R_{H} = \rho_{H}/H$ at $H$ = 10 kOe and 90 kOe.}
\label{FIG6}
\end{figure}

In Fig. \ref{FIG6} the Hall resistivity, $\rho_H$, curves of Re${_3}$Ge${_7}$ as a function of field are plotted at selected temperatures, where $\rho_H$ is negative for the entire temperature range. We note that the Hall resistivity sample is prepared from the batch grown by cooling at a rate of 2.5 $^{\circ}$C/hr. Above $T_c$, $\rho_H$ is quasi-linear in field and weakly depends on temperature as shown in Fig. \ref{FIG6} (b). Below $T_c$, $\rho_H$ is strongly dependent on temperature. A deviation from the linear field dependence of $\rho_H$ is observed in the temperature range 35$-$50 K as shown in Fig. \ref{FIG6} (c), which indicates the presence of both electron and hole charge carriers. The Hall coefficient $R_{H} = \rho_{H}/H$ curves for $H$ = 10 and 90 kOe are plotted in Fig. \ref{FIG6} (d). Cooling the sample from 300 K, $R_H$ is effectively temperature-independent down to $T_{c}$, then drops sharply at $T_{c}$ and levels off after going through a minimum at $\sim$ 25~K. In the one band model approximation, the carrier density can be estimated from the Hall coefficient $R_{H} = 1/ne$, where $n$ is the carrier density and $e$ is the electron charge. The carrier density is approximated to be around 0.04 per f.~u. at 300 K and $1.5 \times 10^{-4}$ per f.~u. at 2 K. The carrier density drops roughly two orders of magnitude below $T_{c}$. Since $R_H$ is negative over the whole temperature range measured, the carriers are predominantly electrons.

\section{Discussion}

Thermodynamic and transport measurements for Re$_{3}$Ge$_{7}$ indicate an anomaly at $T_{c}$ without any hysteresis, providing strong evidence that a phase transition is occurring below this temperature. Since the anomaly is not sensitive to the magnetic field, the origin of the anomaly can be related to the electronic structure change. A similar resistivity behavior has been observed in the $d$-electron oxides, where the transition is driven by strong electron-electron interactions.\cite{Imada1998} In these metallic systems, the structural deformation is often accompanied by localization of carriers, causing a metal-insulator transition or a transition showing a sudden increase in resistivity on cooling across the phase transition. The ground state below the phase transition could be a spin-, charge- or orbital-ordered state, depending on the lattice and the electronic structure of the system.\cite{Imada1998} However, based on the magnetization and specific heat results, strong electron-electron interactions are not expected for the Re$_{3}$Ge$_{7}$ compound. In general, the specific heat shows sharp $\delta$-function-like behavior (first order phase transition) at the structural phase transition. However, the specific heat of Re$_{3}$Ge$_{7}$ shows a $\lambda$-like anomaly at $T_c$. The crystal structure of Re$_{3}$Ge$_{7}$ can be described by stacking distorted hexagonal planes along the $b$-direction, consisting of isolated ReGe$_{3}$ prisms and double Re$_{2}$Ge$_{4}$ prisms.\cite{Siegrist1983} Thus, the structure consists of alternating chains of isolated prisms and double prisms, where the double prisms are elongated to locate Re pairs. As the temperature decreases, these alternating prisms can be distorted further to provide space for the Re atoms, which may induce a structural phase transition below $T_{c}$ and localize the carriers into the prisms. Thus, it is necessary to perform X-ray diffraction measurement below $T_{c}$ to check whether there is a structural transition.

\begin{figure}
\centering 
\includegraphics[width=1\linewidth]{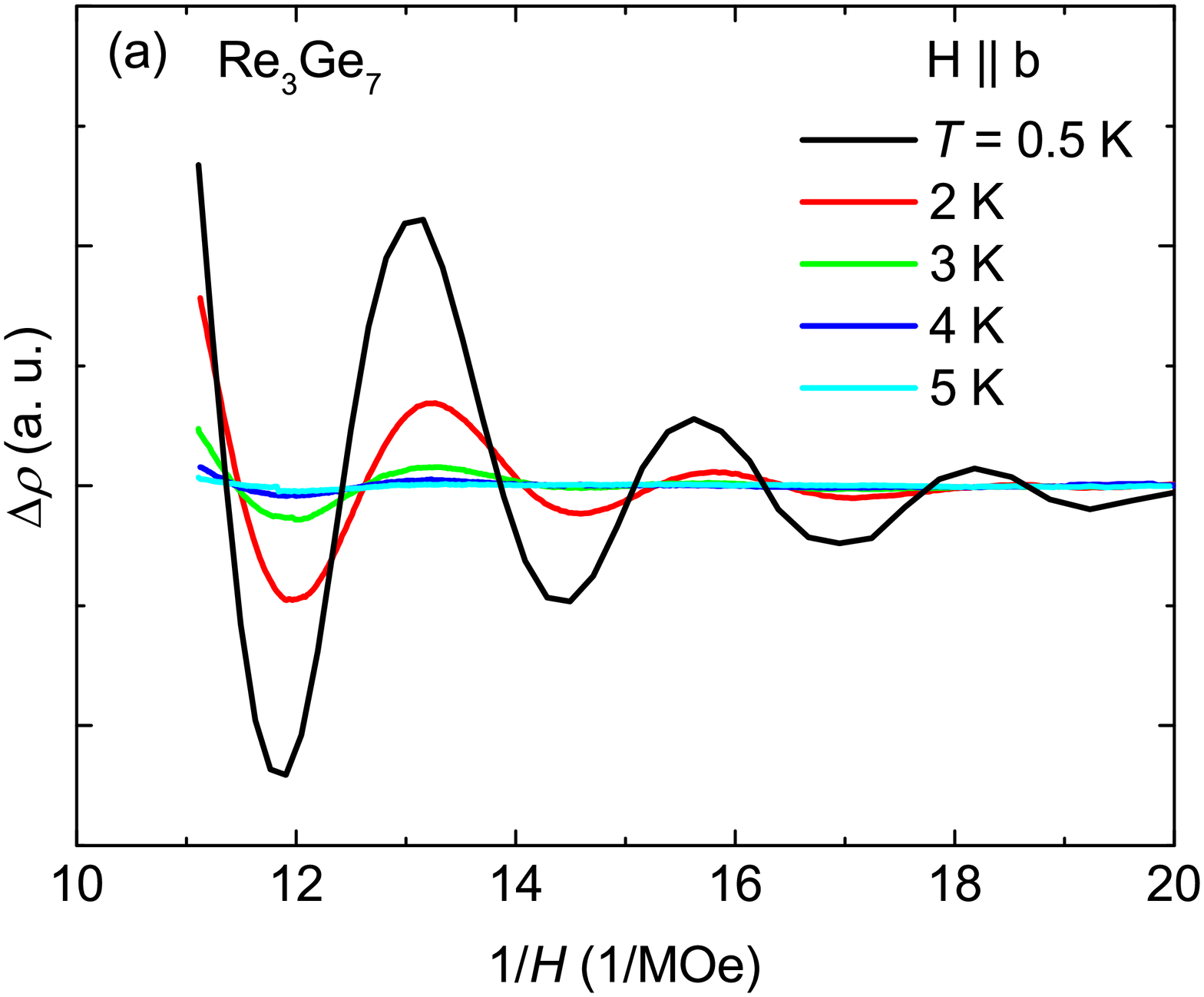}
\includegraphics[width=1\linewidth]{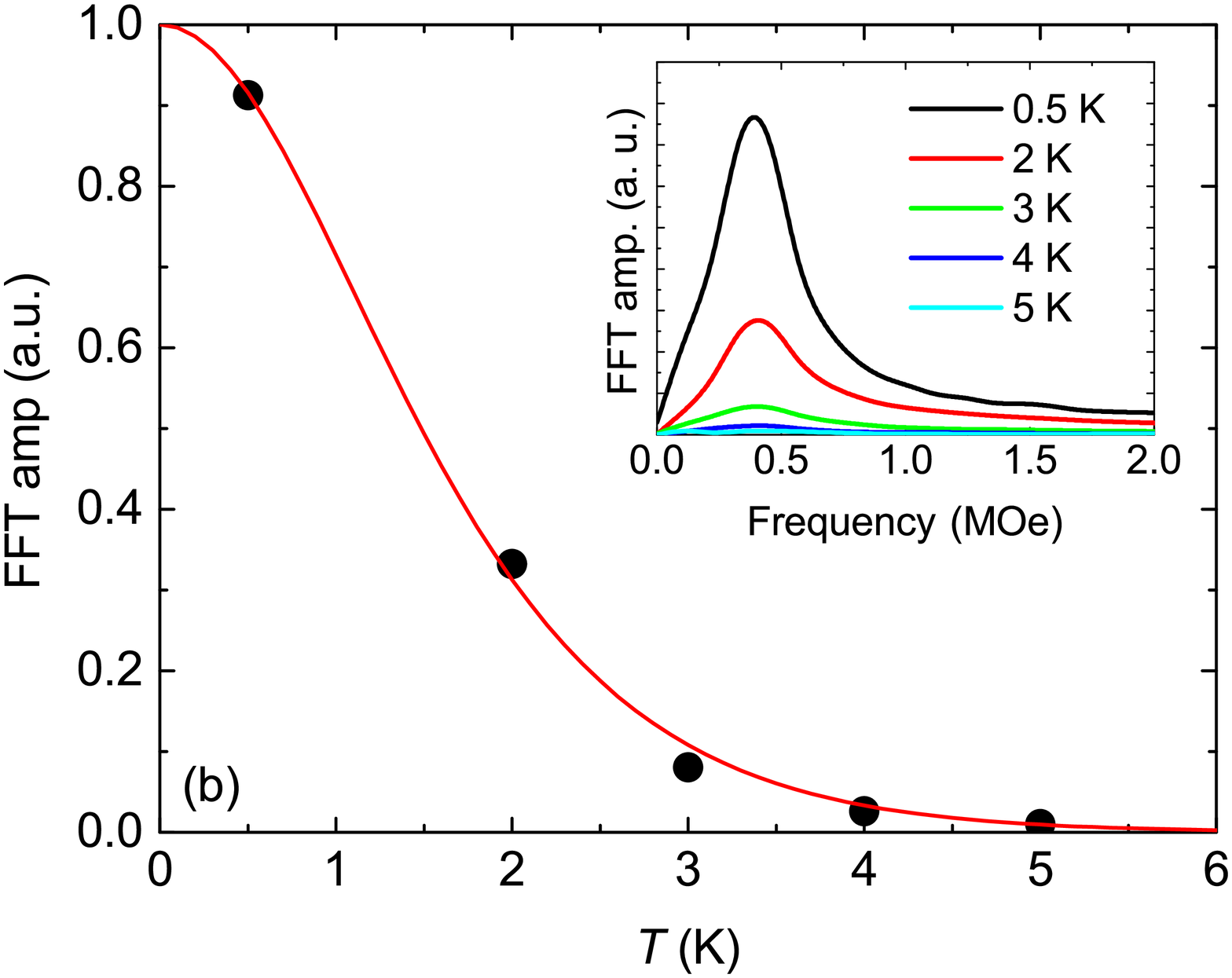}
\caption{(a) Resistivity of Re$_{3}$Ge$_{7}$ at $T$ = 0.5, 2, 3, 4, and 5 K, plotted after subtracting background contributions. (b) Temperature dependence of the SdH amplitude. Solid line represents the fit to the Lifshitz-Kosevich formula. Inset shows FFT spectra of SdH data.}
\label{FIG7}
\end{figure}

The MR of Re${_3}$Ge${_7}$ reveals quantum oscillations in the high magnetic field regime up to 5 K. After subtracting the background MR by using a polynomial fit, the residual oscillations are plotted against $1/H$ in Fig.~\ref{FIG7}~(a). The amplitude of the oscillations decreases with increasing temperature and become too small to discern at 5~K. The fast Fourier transform (FFT) confirms that the oscillations are periodic in 1/$H$ and reveals a single frequency $f$~=~0.45~MOe as shown in the inset of Fig.~\ref{FIG7}~(b). The low frequency of quantum oscillations indicates the small Fermi surface. It is necessary to perform an angular dependence of the SdH measurements to unambiguously determine the Fermi surface topology. The effective mass $m^{*}$ of the carriers can be determined by fitting the temperature-dependent amplitude of the oscillations to the Lifshitz-Kosevich formula\cite{Shoenberg1984}, shown in Fig.~\ref{FIG7}~(b). The fitted effective mass is $m^{*} \sim 0.7~m_{e}$, where $m_{e}$ is the bare electron mass. Since the obtained effective mass is close to the bare electron mass, the small $\gamma$ value of Re$_{3}$Ge$_{7}$ is probably due to the low carrier density. Thus, the specific heat of Re$_{3}$Ge$_{7}$ is dominated by phonon contributions at low temperatures, behaving as the insulator $C_{p} \propto T^{3}$. This conjecture is supported by the transport measurements, where the resistivity shows a metal-to-insulator-like transition and the Hall coefficient measurement indicates a reduction of the carrier density by two orders of magnitude below $T_{c}$.

\begin{figure}
\centering 
\includegraphics[width=1\linewidth]{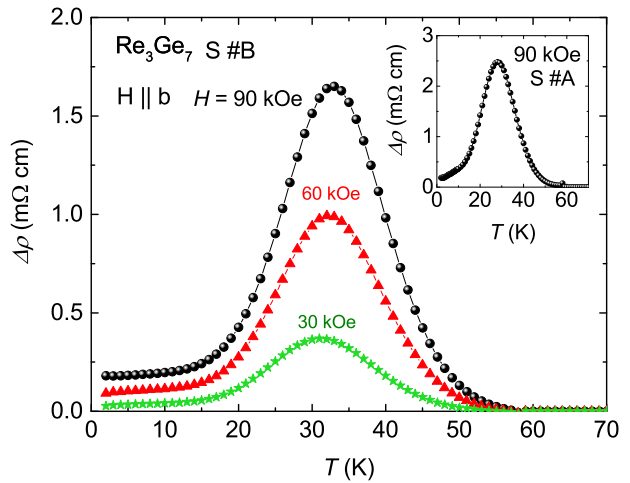}
\caption{Temperature dependence of the magnetoresistance $\Delta\rho = \rho(T, H) - \rho(T, H = 0)$ curves of Sample B for $H$ = 30, 60 and 90 kOe. Inset shows $\Delta\rho$ of Sample A for $H$ = 90 kOe.}
\label{FIG8}
\end{figure}

The resistivity of Re${_3}$Ge${_7}$ single crystals follows ordinary metallic behavior above $T_c$, whereas a broad maximum or shoulder develops in resistivity below $T_{c}$. Although the apparently different resistivity behavior below $T_{c}$ is observed for Re$_{3}$Ge$_{7}$ single crystals, a closer examination of the resistivity curves indicates that the characteristic features in $\rho(T)$ actually take place at nearly the same temperature. This implies that this anomaly might have a common origin that is intrinsic to Re$_{3}$Ge$_{7}$. In addition, the similar effects of magnetic fields on the resistivity, shown in Fig. \ref{FIG4}, further support this. To clarify the effect of applying a magnetic field on the resistivity, the zero field resistivity curve for Sample B is subtracted from the $H$ = 30, 60 and 90 kOe curves and the resulting $\Delta\rho = \rho(T, H) - \rho(T, H = 0)$ curves are plotted in Fig.~\ref{FIG8}. As shown in the figure, the effect of the magnetic field on $\rho(T)$ is anomalous, deviating from ordinary metallic behavior, displaying nearly symmetric maxima around 30~K. The $\Delta\rho (T)$ curve for an ordinary metal would be expected to increase with decreasing temperature, then level off at low temperatures. However, the $\Delta\rho (T)$ of Re$_3$Ge$_7$ steadily increases as temperature is lowered below $T_c$ and unexpectedly goes through a maximum at $\sim$30~K. The height of the maximum is strongly dependent on the strength of the applied magnetic field. In addition, the maximum position shifts slightly towards a higher temperature with increasing magnetic fields. It should be noted that $\Delta\rho(T)$ above $T_{c}$ exhibits no obvious enhancement (Fig. \ref{FIG5}). The $\Delta\rho (T)$ curve of Sample~A, plotted in the inset of Fig. \ref{FIG8} for $H$ = 90 kOe, shows similar behavior. Though the absolute value of $\Delta\rho(T)$ is different for different samples, the temperature dependence shows the same general behavior.

A similar resistivity behavior under applied magnetic fields has been observed in several compounds such as LaSb$_{2}$\cite{Budko1998}, $R$PtBi ($R$ = rare-earth)\cite{Mun2016}, silver chalcogenides\cite{Husmann1997,Lee2002}, and pentatellurides\cite{DiSalvo1981}. In zero field, highly anisotropic LaSb$_{2}$ compound shows typical metallic behavior and reveals no magnetic ordering or structural phase transition below 300 K. The resistivity develops a broad peak around $\sim$12~K for $H$~=~25~kOe. As the magnetic field increases, the broad peak position moves towards higher temperatures, forming a peak at $\sim$23~K for $H$~=~180~kOe.\cite{Budko1998} It has been suggested that the anomaly induced by the magnetic field is due to a magnetic-field-induced density wave transition.\cite{Chaikin1985} The field induced resistivity maximum is also observed in $R$PtBi (R = rare-earth) compounds \cite{Mun2016} and silver chalcogenides \cite{Husmann1997,Lee2002}, where the resistivity maximum is observed at the sign change of the Hall coefficient and thermoelectric power due to the band crossing. 
Note that for $R$PtBi compounds, a large and nearly linear MR is observed for both the cases with sign reversal and without sign reversal in thermoelectric power and Hall coefficient.\cite{Mun2016} Interestingly, Re$_{3}$Ge$_{7}$ also shows a linear MR in the high field regime without sign change in the Hall coefficient. Layered pentatellurides ZrTe$_{5}$ and HfTe$_{5}$ with one-dimensional transition-metal chains show a peak in the resistivity. In zero field the maximum is located at $\sim$ 130 K for ZrTe$_{5}$\cite{Li2016} and $\sim$ 65 K for HfTe$_{5}$\cite{Lv2018} compound without any signature of structural distortion such as charge density wave \cite{Okada1982}. As magnetic field increases, the peak position shifts toward higher temperature. A change in sign of the thermoelectric power at the resistivity maximum is also observed for ZrTe$_{5}$.\cite{Shahi2018} The resistivity anomaly observed in pentatellurides has been explained by multiple scenarios, including semimetal to semiconductor transition\cite{McIlroy2004}, temperature induced Lifshitz transition\cite{Zhang2017}, or bipolar conduction mechanism\cite{Shahi2018}. Recently, these compounds have become of interest as topologically non-trivial materials. It is noteworthy that different classes of materials share qualitatively similar behavior but that the mechanism of forming the resistivity maximum may differ. For Re$_{3}$Ge$_{7}$, the origin of the anomalous resistivity is currently unknown. Further studies are necessary to explain multiple phenomena observed in Re$_{3}$Ge$_{7}$: i) the metal-to-insulator-like phase transition below $T_{c}$ = 58.5 K, ii) the resistivity anomaly around 30 K, iii) the field induced resistivity maximum around 30 K, iv) the negative MR in low field regime at low temperatures, and v) the linear field dependence of MR in high field regime. Synchrotron X-ray and angle resolved photoemission spectroscopy (ARPES) measurements and resistivity measurements under hydrostatic pressure are under way to clarify the observed anomalous transport properties of Re$_{3}$Ge$_{7}$.

\section{Summary}

Single crystals of Re${_3}$Ge${_7}$ were synthesized from high temperature melt and characterized by measuring magnetization, electrical resistivity, Hall coefficient, and specific heat. Based on measurements of thermodynamic and transport properties, Re${_3}$Ge${_7}$ can be classified as a weakly diamagnetic, intermetallic compound with a phase transition below $T_{c}$ = 58.5 K. In zero field, a drastic increase in resistivity below $T_{c}$ indicates a metal-to-insulator-like transition. The application of the magnetic field induces an anomalous resistivity behavior, deviating from that of normal metallic behavior, included a magnetic field induced resistivity maximum around 30~K, the negative MR in low fields, and the linear MR in high fields. The carrier concentration inferred from the Hall coefficient measurement decreases two order of magnitude below $T_{c}$. The effective mass of charge carriers, inferred from Shubnikov-de Haas quantum oscillations, is close to the bare electron mass.

\section{Acknowledgements}
This work was supported by the Canada Research Chairs program, the Natural Science and Engineering Research Council of Canada, the Canadian Institute for Advanced Research, and the Canadian Foundation for Innovation.

\end{document}